\documentclass[prc,preprint]{revtex4}

\usepackage{graphicx}
\usepackage{amsmath}
\begin{document}

\title{{\bf Orbital angular momentum transfer via spontaneously generated coherence}}

\author{{\bf Zahra Amini Sabegh, Mohammad Mohammadi, Mohammad Ali Maleki and Mohammad Mahmoudi \footnote{E-mail: mahmoudi@znu.ac.ir}}}
 \affiliation{Department of Physics, University of Zanjan, University Blvd., 45371-38791, Zanjan, Iran}

\begin{abstract}
We study the orbital angular momentum (OAM) transfer from a weak Laguerre-Gaussian (LG) field to a weak plane-wave in two closed-loop three-level $V$-type atomic systems. In the first scheme, the atomic system has two non-degenerate upper levels which the corresponding transition is excited by a microwave plane-wave. It is analytically shown that the microwave field induces an OAM transfer from an LG field to a generated third field. In the second scheme, we consider a three-level $V$-type atomic system with two near-degenerate excited states and study the effect of the quantum interference due to the spontaneous emission on the OAM transfer. It is found that the spontaneously generated coherence (SGC) induces the OAM transfer from the LG field to the weak planar field, while the OAM transfer does not occur in the absence of the SGC. The suggested models prepare a rather simple method for the OAM transfer which can be used in quantum information processing and data storage.
\end{abstract}
\maketitle

\section{Introduction}
Quantum interference has a major role in determining the optical properties of the quantum systems \cite{ficek}. It is well-known that the spontaneously generated coherence (SGC) is a kind of vacuum induced coherence which may give rise to a coherent superposition of two quantum states. The vacuum induced coherence can be occurred in $\Lambda$ \cite{javanainen1992}, $V$ \cite{Paspalakis1998} or ladder type \cite{ladder2004} quantum systems. In the past two decades, the effect of the vacuum induced coherence on the optical properties of the quantum system has been studied and it was shown that the SGC can affect the optical phenomena such as pump-probe response \cite{menon1998}, loss-free propagation of a short laser pulse \cite{Paspalakis1999}, Autler-Townes doublet \cite{menon1999}, optical bistability \cite{joshi2003}, spectral-line narrowing \cite{PRA2008}, Kerr nonlinearity \cite{PLA2008,commu2009}, electromagnetically induced transparency (EIT), and refractive properties \cite{EPJD2009}. The role of quantum interference in the optical properties beyond two-photon resonance condition is investigated in a three-level V-type pump–probe atomic system and it was demonstrated that the absorption, dispersion, group index, and optical bistability depend on the SGC \cite{Prof}.

On the other hand, in 1992, Allen \textit{et al.} have found that a laser beam can carry orbital angular momentum (OAM), so-called the vortex beam, in addition to the spin angular momentum of the circularly polarized beam \cite{PRA1992}. A fine particle rotates around its central axis when it is exerted by a polarized light carrying spin angular momentum. However, a light beam with OAM makes the particle to orbit in some special conditions and the OAM of light transfers to absorptive particles \cite{rotation1995}. In recent years, the vortex beams have been taken into consideration because of their potential to create the high-dimensional Hilbert space. One of the vortex beams is a Laguerre-Gaussian (LG) beam which can be generated using forked diffraction gratings \cite{forked1990}, cylindrical lens mode converters \cite{HG1993}, spiral phase plates \cite{plate1994}, holograms \cite{holo1992}, and spatial light modulators (SLMs) \cite{SLM2007}. An LG beam has a doughnut-like intensity profile with helical wavefront due to the azimuthal component of its Poynting vector. Some of the quantum optical phenomena such as EIT \cite{PRL2015}, four-wave mixing (FWM) \cite{PRL2012,lett2004,PRB2015,PRA2017}, second harmonic generation (SHG) \cite{PRA1996,PRA1997}, sum frequency generation \cite{JOSAB2015}, and atom-photon entanglement \cite{Srep2018} have been controlled by the OAM of light. Recently, the OAM transfer from a vortex beam to a non-vortex beam via the light-matter interaction has been widely investigated in different atomic systems \cite{PRA2001,PRA2011,PRA2013,Hamedi2018,Srep2019}.

In this manuscript, we are going to study the possibility of the OAM transfer via quantum interference due to the spontaneous emission which has not been previously reported. The OAM transfer between two laser beams is investigated in two simple three-level atomic systems with non-degenerate and near-degenerate two upper levels. Analytical solution of the Bloch and coupled Maxwell equations gives the Rabi frequency of the output field along the propagation direction. It is found that the OAM transfer between light beams depends on the microwave field in non-degenerate excited states as well as the strength of the SGC in near-degenerate states. The absorption profiles for different modes of the LG field show that the absorption(gain)-assisted EIT window in the atomic system leads to enhancement of the OAM transfer efficiency. The OAM transfer in these simple atomic systems can be used for broad applications in quantum information processing. We organize this manuscript as follows. Section 2 is devoted to the OAM transfer in a closed-loop three-level $V$-type atomic system. In section 3, we introduce a three-level $V$-type atomic system and show the OAM transfer induced by the SGC. Section 4 presents our conclusions.


\section{Closed-Loop Three-Level $V$-Type Atomic Systems}

\begin{figure}[htbp]
\centering
  \includegraphics[width=7.0cm]{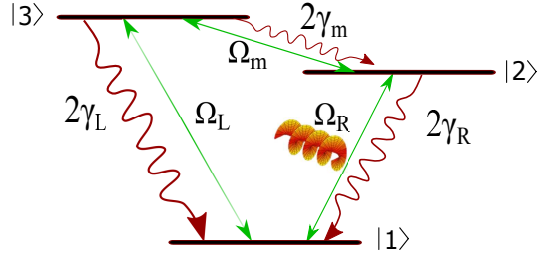}\\
  \caption{\small Closed-loop three-level $V$-type atomic system schematics which can be established in Rb Rydberg vapor sample interacting with two weak applied fields, $\Omega_{R}$ and $\Omega_{L}$ and a strong planar microwave field, $\Omega_{m}$. The spontaneous emission rates are defined by $\gamma_{R}$, $\gamma_{L}$, and $\gamma_{m}$.}\label{f1}
\end{figure}

Let us consider an ensemble of the closed-loop three-level $V$-type atoms with two non-degenerate excited states, as shown in Fig. \ref{f1}. Two weak probe fields, named as right and left fields, are applied to the $|2\rangle\leftrightarrow|1\rangle$ and $|3\rangle\leftrightarrow|1\rangle$ transitions, respectively. The transition $|3\rangle\leftrightarrow|2\rangle$ is driven by a planar microwave field. The Rabi frequency of the applied fields is defined by $\Omega_{\sigma}=\vec{\mu}_{\sigma}\cdot\vec{E}_{\sigma}/\hbar$ with $\sigma=R$, $L$, and $m$. The atomic induced dipole moment of the related transition and amplitude of the field vectors are denoted by $\vec{\mu}_{\sigma}$ and $\vec{E}_{\sigma}$, respectively. The proposed atomic system can be established in the Rb Rydberg atoms \cite{Rydberg}. The spontaneous emission rates are indicated by $\gamma_{R}$, $\gamma_{L}$, and $\gamma_{m}$.

The right field is chosen to be a planar field and the left planar field is generated along the atomic medium due to the microwave field. We suppose that the right field is an LG one which carries the OAM and its Rabi frequency in the cylindrical coordinates is given by
\begin{equation}\label{e1}
    \Omega_{R}(r,\varphi)=\Omega_{0R}\frac{1}{\sqrt{|l|!}}(\frac{\sqrt{2}r}{w_{LG}})^{|l|}~e^{-r^{2}/w_{LG}^{2}}~e^{il\varphi}.
\end{equation}
Here, $\Omega_{0R}$, $w_{LG}$, and $l$ describe the constant Rabi frequency of the right field, LG beam waist, and OAM index, respectively. In order to investigate the properties of the generated left field, it is necessary to solve the Bloch and Maxwell equations, simultaneously. The Bloch equations are obtained using density matrix formalism in the electric-dipole and rotating-wave approximations as
\begin{eqnarray}\label{e2}
\dot{\rho}_{22}&=&-2\gamma_{R}\rho_{22}+2\gamma_{m}\rho_{33}+i\Omega^{\ast}_{R}\rho_{12}-i\Omega_{R}\rho_{21}+i\Omega_{m}\rho_{32}-i\Omega^{\ast}_{m}\rho_{23},\nonumber\\
\dot{\rho}_{33}&=&-2(\gamma_{L}+\gamma_{m})\rho_{33}+i\Omega^{\ast}_{L}\rho_{13}-i\Omega_{L}\rho_{31}-i\Omega_{m}\rho_{32}+i\Omega^{\ast}_{m}\rho_{23},\nonumber\\
\dot{\rho}_{12}&=&(i\Delta_{R}-\gamma_{R})\rho_{12}-i\Omega_{R}(\rho_{22}-\rho_{11})+i\Omega_{L}\rho_{32}-i\Omega^{\ast}_{m}\rho_{13},\nonumber\\
\dot{\rho}_{13}&=&[i\Delta_{L}-(\gamma_{L}+\gamma_{m})]\rho_{13}-i\Omega_{L}(\rho_{33}-\rho_{11})+i\Omega_{R}\rho_{23}-i\Omega_{m}\rho_{12},\nonumber\\
\dot{\rho}_{23}&=&[i\Delta_{m}-(\gamma_{R}+\gamma_{L}+\gamma_{m})]\rho_{23}-i\Omega_{L}\rho_{21}+i\Omega^{\ast}_{R}\rho_{13}+i\Omega_{m}(\rho_{22}-\rho_{33}),\nonumber\\
\dot{\rho}_{11}&=&-(\dot{\rho}_{22}+\dot{\rho}_{33}),
\end{eqnarray}
where the frequency detuning between the laser field and corresponding transition is indicated by  $\Delta_{\sigma}=\omega_{\sigma}-\bar{\omega}_{\sigma}$. The analytical expressions for the coherence terms, $\rho_{21}$ and $\rho_{31}$, can be obtained solving Eq. (\ref{e2}) in the steady state and under multi-photon resonance condition, $\Delta_{R}=\Delta_{L}=\Delta_{m}=0$, $\gamma_{R}=\gamma_{L}=\gamma$, and $\gamma_{m}=2\gamma$, as
\begin{eqnarray}\label{e3}
  \rho_{21}&=&\frac{-\Omega_{L}\Omega_{m}-3i\gamma\Omega_{R}}{3\gamma^{2}+\Omega_{m}^{2}},\nonumber\\
  \rho_{31}&=&\frac{i\gamma\Omega_{L}-\Omega_{m}\Omega_{R}}{3\gamma^{2}+\Omega_{m}^{2}}.
\end{eqnarray}
We consider that the positive(negative) imaginary part of the coherence terms shows the absorption(gain) response of the atomic system.
On the other hand, the Maxwell equations for the weak time-independent right and left fields, propagating in the $z$-direction, in the slowly varying envelope approximation stand for
\begin{eqnarray}\label{e4}
  \frac{\partial\Omega_{R}(z)}{\partial z}&=&i\frac{\alpha\gamma}{2L}\rho_{21},\nonumber\\
  \frac{\partial\Omega_{L}(z)}{\partial z}&=&i\frac{\alpha\gamma}{2L}\rho_{31},
\end{eqnarray}
where $\gamma_R=\gamma_L=\gamma$.  The parameter $L$ is the length of the atomic medium and $\alpha$ defines the optical depth for both right and left fields \cite{Hamedi2018}. Substituting Eq. (\ref{e3}) in Eq. (\ref{e4}) and assuming $\Omega_{L}(z=0)=0$ and $\Omega_{R}(z=0)=\Omega_{R}(r,\varphi)$, given by Eq. (\ref{e1}), the Rabi frequency of the generated left field in the atomic medium takes the form
\begin{eqnarray}\label{e5}
  \Omega_{L}(z)&=&\frac{i}{2\sqrt{\gamma^{2}-\Omega_{m}^{2}}}[\exp(\frac{-z\alpha\gamma(2\gamma+\sqrt{\gamma^{2}-\Omega_{m}^{2}})}{2L(3\gamma^{2}+\Omega_{m}^{2})})\nonumber\\
  &-&\exp(
  \frac{z\alpha\gamma(-2\gamma+\sqrt{\gamma^{2}-\Omega_{m}^{2}})}{2L(3\gamma^{2}+\Omega_{m}^{2})})]\Omega_{m}\Omega_{R}(r,\varphi).
\end{eqnarray}
\begin{figure}[t!]
\centering
  \includegraphics[width=0.7\textwidth]{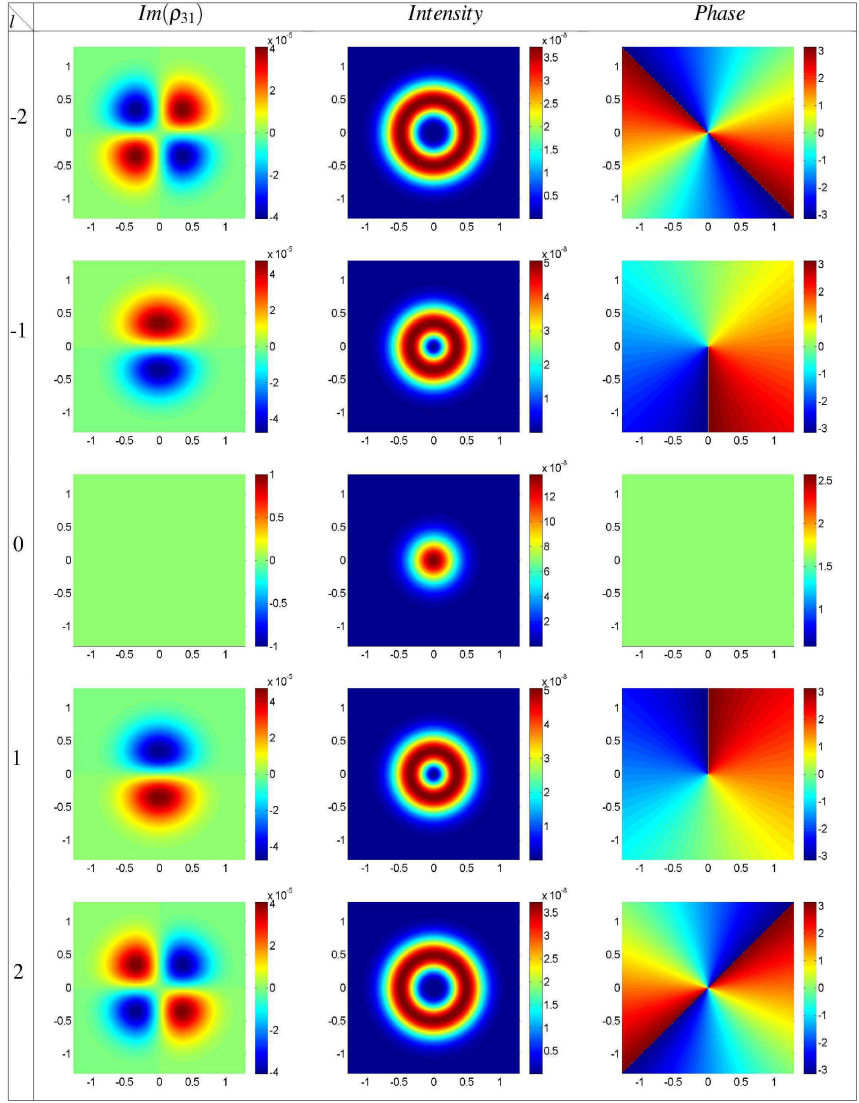}
  \caption{\small The spatially dependent behavior of the imaginary part of $\rho_{31}$ (left column), intensity (middle column), and phase (right column) profiles of the output left field versus $x$ and $y$ for the Gaussian and different modes of the right LG field, i.e., $l=-2,-1,...,2$, at $z=L$. Used parameters are $\gamma_{R}=\gamma_{L}=\gamma$, $\gamma_{m}=2\gamma$, $\Omega_m=4\gamma$, $\Omega_{0R}=0.1\gamma$, $w_G=w_{LG}=0.5 mm$, $\alpha=100$, and $\Delta_{R}=\Delta_{L}=\Delta_{m}=0$.}\label{f2}
\end{figure}
It is found that the Rabi frequency of the generated left field corresponds to the Rabi frequency of the right field. It means that by choosing the right field as an LG field, the OAM transfers to the generated left field. It is worth to note that, here, the microwave field is directly responsible for the OAM transfer from the right to left field. According to Eq. (\ref{e5}), the left field cannot be generated in the absence of the microwave field, so the OAM transfer is stopped.

In the following, we are going to study the phase and intensity properties of the output left field. We scale all the frequencies by $\gamma$. Using Eqs. (\ref{e3}) and (\ref{e5}), we plot the absorption (left column), intensity (middle column), and phase (right column) profiles of the generated left field versus $x$ and $y$ at the end plane of the atomic medium, $z=L$, for the Gaussian and different modes of the right LG field, i.e., $l=-2,-1,...,2$, in Fig. \ref{f2}. The horizontal and vertical axes are $x$ and $y$ axes in millimeter. Used parameters are $\gamma_{R}=\gamma_{L}=\gamma$, $\gamma_{m}=2\gamma$, $\Omega_m=4\gamma$, $\Omega_{0R}=0.1\gamma$, $w_G=w_{LG}=0.5 mm$, $\alpha=100$, and $\Delta_{R}=\Delta_{L}=\Delta_{m}=0$. As expected from Eq. (\ref{e5}), the output left field is directly proportional to the microwave and right fields. The generation of the left field depends on the increment of the medium transparency at the end plane of the atomic medium. The absorption profiles, in Fig. \ref{f2}, indicate that the medium is nearly transparent at the end plane of the atomic medium. A bird's eye view on the results of Fig. \ref{f2} shows that the number of petals in the absorption profiles is $2l$ which indicate the absorption or gain regions. An investigation on the middle and right columns shows that the output left field contains the properties of an LG field and carries a finite OAM. For the special case, when the right field is considered as a Gaussian with the planar wavefront, the output left field becomes Gaussian with uniform spatially phase profile distribution.

\section{Three-Level $V$-Type Atomic Systems with the SGC}

\begin{figure}[htbp]
\centering
  \includegraphics[width=7.0cm]{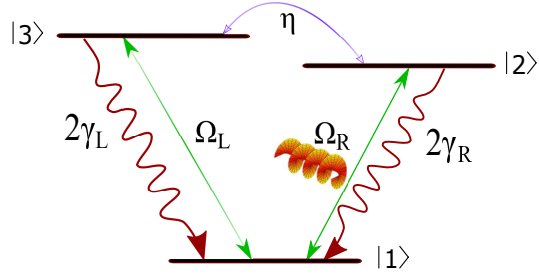}\\
  \caption{\small Three-level $V$-type atomic system schematics which can be established in a sodium vapor sample with the SGC interacting with two weak applied fields, $\Omega_{R}$ and $\Omega_{L}$. The spontaneous emission rates from two upper energy levels to the ground state are defined by $\gamma_{R}$ and $\gamma_{L}$.}\label{f3}
\end{figure}

 In order to study the OAM transfer via the SGC, we consider an ensemble of three-level $V$-type atoms including two near-degenerate upper levels as shown in Fig. \ref {f3} and take into account the effect of the quantum interference due to the SGC. The strength of the SGC is given by $\eta=\vec{\mu}_{31}\cdot\vec{\mu}_{21}/|\vec{\mu}_{31}|\cdot|\vec{\mu}_{21}|$ which depends on the angle between two atomic induced dipole moment vectors. It is assumed that the transition $|3\rangle\leftrightarrow|1\rangle$($|2\rangle\leftrightarrow|1\rangle$) is excited by left(right) field with Rabi frequency $\Omega_{L}=\vec{\mu}_{31}\cdot\vec{E}_{L}/\hbar$($\Omega_{R}=\vec{\mu}_{21}\cdot\vec{E}_{R}/\hbar$) and laser frequency $\omega_{L}$($\omega_{R}$). The atomic induced dipole moment of the transition $|3\rangle\leftrightarrow|1\rangle$($|2\rangle\leftrightarrow|1\rangle$) and amplitude of left(right) field are denoted by $\vec{\mu}_{31}$($\vec{\mu}_{21}$) and $\vec{E}_{L}$($\vec{E}_{R}$), respectively. The spontaneous emission rates from two upper energy levels to the ground state are defined by $\gamma_{R}$ and $\gamma_{L}$. As a  realistic example, we consider $D_{2}$ line of sodium atoms with a ground state $|1\rangle=|3~^{2}S_{1/2},F=1\rangle$, two sublevels $|3~^{2}P_{3/2},F'=0\rangle$ and $|3~^{2}P_{3/2},F'=1\rangle$ as two excited states, $|2\rangle$ and $|3\rangle$, respectively \cite{sodium}.

The Bloch equations in the electric-dipole and rotating-wave approximations, read to
\begin{eqnarray}\label{e6}
\dot{\rho}_{22}&=&-2\gamma_{R}\rho_{22}+i\Omega^{\ast}_{R}\rho_{12}-i\Omega_{R}\rho_{21}-\eta \sqrt{\gamma_{R}\gamma_{L}}(\rho_{32}+\rho_{23}),\nonumber\\
\dot{\rho}_{33}&=&-2\gamma_{L}\rho_{33}+i\Omega^{\ast}_{L}\rho_{13}-i\Omega_{L}\rho_{31}-\eta \sqrt{\gamma_{R}\gamma_{L}}(\rho_{32}+\rho_{23}),\nonumber\\
\dot{\rho}_{12}&=&(i\Delta_{R}-\gamma_{R})\rho_{12}-i\Omega_{R}(\rho_{22}-\rho_{11})+i\Omega_{L}\rho_{32}-\eta \sqrt{\gamma_{R}\gamma_{L}}\rho_{13},\nonumber\\
\dot{\rho}_{13}&=&(i\Delta_{L}-\gamma_{L})\rho_{13}-i\Omega_{L}(\rho_{33}-\rho_{11})+i\Omega_{R}\rho_{23}-\eta \sqrt{\gamma_{R}\gamma_{L}}\rho_{12},\nonumber\\
\dot{\rho}_{23}&=&[i(\Delta_{L}-\Delta_{R})-(\gamma_{R}+\gamma_{L})]\rho_{23}-i\Omega_{L}\rho_{12}+i\Omega^{\ast}_{R}\rho_{13}-\eta \sqrt{\gamma_{R}\gamma_{L}}(\rho_{22}+\rho_{33}),\nonumber\\
\dot{\rho}_{11}&=&-(\dot{\rho}_{22}+\dot{\rho}_{33}).
\end{eqnarray}
We assume $\gamma_{R}=\gamma_{L}$ and scale all the frequencies by $\gamma=2\pi\times9.8 MHz$. Solving Eq. (\ref{e6}) for $\Delta_{R}=\Delta_{L}=0$ gives the coherence terms of the probe transitions as \cite{Prof}
\begin{eqnarray}\label{e7}
  \rho_{21}&=&i\frac{\eta(\eta-1)^{2}\Omega_{L}+(\eta-1)^{2}\Omega_{R}}{3\eta^{2}(1-\eta^{2})-1},\nonumber\\
  \rho_{31}&=&i\frac{\eta(\eta-1)^{2}\Omega_{R}+(\eta-1)^{2}\Omega_{L}}{3\eta^{2}(1-\eta^{2})-1}.
\end{eqnarray}
We solve the Maxwell equations for the right and left fields in slowly varying approximation using Eq. (\ref{e7}) for the selected boundary conditions, $\Omega_{L}(z=0)=0$ and $\Omega_{R}(z=0)=\Omega_{R}(r,\varphi)$. We obtain the Rabi frequency of the generated left field, propagating in the $z$-direction, as
\begin{eqnarray}\label{e8}
  \Omega_{L}(z)&=&-\frac{1}{2}[\exp(\frac{z\alpha(-\eta-1)(1-2\eta^{2}+\eta^{4})}{2L(1-3\eta^{2}+3\eta^{4})})\nonumber \\
  &-&\exp(
  \frac{z\alpha(\eta-1)(1-2\eta^{2}+\eta^{4})}{2L(1-3\eta^{2}+3\eta^{4})})]\Omega_{R}(r,\varphi).
\end{eqnarray}
Two remarks are in order about result \ref{e8}. First, in the absence of the SGC, the amplitude of the generated left field vanishes and the OAM transfer does not occur. So, the presence of the SGC leads to the OAM transfer from the right field to the left one. Second, the properties of the generated left field directly depend on the intensity and phase of the right LG field. Thus, the OAM transfer in the three-level $V$-type atomic systems with two near-degenerate excited states is due to the SGC.

\begin{figure}[t!]
\centering
  \includegraphics[width=0.7\textwidth]{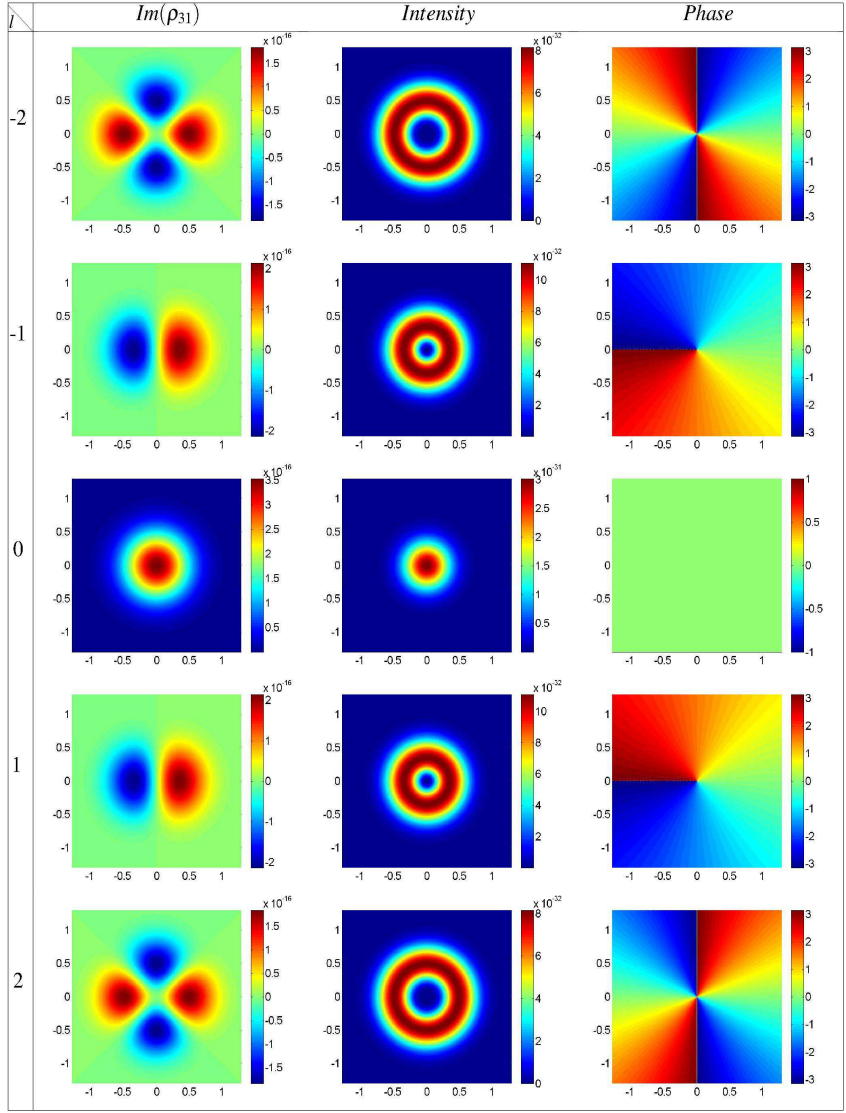}
  \caption{\small The spatially dependent behavior of the imaginary part of $\rho_{31}$ (left column), intensity (middle column), and phase (right column) profiles of the output left field as a function of $x$ and $y$ for the Gaussian and different modes of the right LG field, i.e., $l=-2,-1,...,2$ at $z=L$. Used parameters are $\eta=0.5$, $\gamma_{R}=\gamma_{L}=\gamma$, $\Omega_{0R}=0.1\gamma$, $w_G=w_{LG}=0.5 mm$, $\alpha=100$, and $\Delta_{L}=\Delta_{R}=0$.}\label{f4}
\end{figure}

Now, we are interested in studying the OAM transfer between light beams in the three-level $V$-type atomic system in the presence of the SGC. The absorption, intensity, and phase profiles of the generated left field at the end plane of the atomic medium, $z=L$, are obtained using Eqs. (\ref{e7}) and (\ref{e8}).  In Fig. \ref{f4}, the spatially dependent behavior of the imaginary part of $\rho_{31}$ (left column), intensity (middle column), and phase (right column) profiles of the output left field as a function of $x$ and $y$ are plotted for the Gaussian and different modes of the right LG field, i.e., $l=-2,-1,...,2$ at $z=L$. Used parameters are $\eta=0.5$, $\gamma_{L}=\gamma_{R}=\gamma$, $\Omega_{0R}=0.1\gamma$, $w_G=w_{LG}=0.5 mm$, $\alpha=100$, and $\Delta_{L}=\Delta_{R}=0$. The intensity and phase profiles of the output left field (middle and right columns) are absolutely equivalent to the different modes of the right LG field. Although the medium becomes more transparent in comparing with Fig. \ref{f2}, the output left field intensity is heavily decreased for all the right LG field modes.

\begin{figure}[t!]
\centering
  \includegraphics[width=0.7\textwidth]{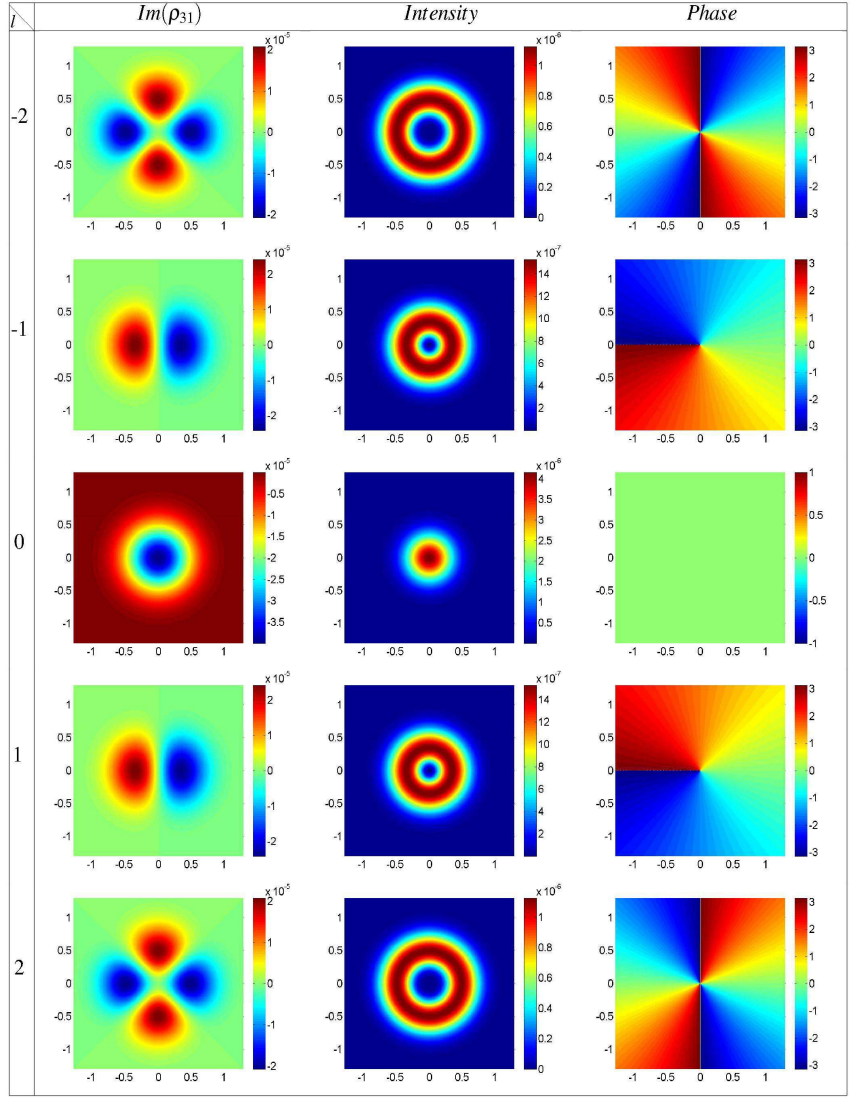}
  \caption{\small The spatially dependent behavior of the imaginary part of $\rho_{31}$ (left column), intensity (middle column), and phase (right column) profiles of the output left field versus $x$ and $y$ for the Gaussian and different modes of the right LG field, i.e., $l=-2,-1,...,2$. The SGC strength is considered to be $\eta=0.99$ and the other parameters are the same parameters as in Fig. \ref{f4}.}\label{f5}
\end{figure}

Let us consider the effect of the strong SGC on the OAM transfer. Figure \ref{f5} shows the different density plots of the output left field for $\eta=0.99$ where other parameters are same as in Fig. \ref{f4}. Comparing the left columns of Figs. \ref{f2} and \ref{f5}, we find that the absorption and gain values at $z=L$ have the same order of magnitude for the strong SGC in the second scheme. An investigation on the middle column of Fig. \ref{f5} shows that the output left field's intensity increases up to ten times of the intensity values in Fig. \ref{f2} for the strong SGC. Therefore, the efficiency of the OAM transfer completely depends on the SGC strength of the atomic system. However, the phase of the right LG field (right column of Figs. \ref{f4} and \ref{f5}) perfectly transfers to the left planar field for nonzero SGC values.

\section{Conclusion}
We have investigated the OAM transfer between two laser beams via the quantum interference due to the SGC. Firstly, the OAM transfer has been studied in a closed-loop three-level $V$-type atomic system with two non-degenerate upper levels. It has been analytically shown that the OAM transfer from the LG field to the planar field is induced by the microwave field, applied to the non-degenerate excited states transition. Secondly, the effect of the SGC on the OAM transfer has been investigated in a three-level $V$-type atomic system with two near-degenerate excited states and has been shown that the OAM transfers in the presence of the SGC. The OAM transfer is accompanied by negligible gain- or absorption-assisted EIT window for different choices of the right LG field's mode. It is worth to note that the left LG field intensity for the strong SGC increases with respect to the first scheme. Controlling the OAM transfer via the SGC can play a crucial role in high-dimensional data storage and data transfer.

\end{document}